\documentclass[10pt,journal]{IEEEtran}
\usepackage{graphicx}          
\usepackage{amsmath}           
\usepackage{amssymb}
\usepackage{algorithm}
\usepackage{algpseudocode}
\usepackage{hyperref}
\usepackage{color}
\usepackage{cite}
\usepackage{enumerate}
\usepackage{epstopdf}
\usepackage{subcaption}
\usepackage[latin1]{inputenc}  

\usepackage{amsthm}

\theoremstyle{definition}

\theoremstyle{remark}
\theoremstyle{plain}

\theoremstyle{remark} 

\DeclareMathOperator{\tr}{tr}
\DeclareMathOperator{\mvec}{vec}
\DeclareMathOperator*{\argmin}{arg\,min}

\newcommand{\mbf}[1]{\mathbf{#1}}
\newcommand{\mbs}[1]{\boldsymbol{#1}}
\newcommand{\what}[1]{\widehat{#1}}

\newcommand{\0}{\mbf{0}}
\newcommand{\n}{\mbf{n}}
\newcommand{\s}{\mbf{s}}
\newcommand{\y}{\mbf{y}}
\newcommand{\e}{\mbf{e}}
\newcommand{\ycov}{\mbf{R}}
\newcommand{\scov}{\mbf{P}}
\newcommand{\A}{\mbf{A}}
\newcommand{\W}{\what{\mbf{W}}}
\newcommand{\I}{\mbf{I}}
\newcommand{\T}{\mbf{T}}
\newcommand{\U}{\what{\mbf{U}}_s}
\newcommand{\Lam}{\mbs{\Lambda}}
\newcommand{\G}{\what{\mbs{\Gamma}}}
\newcommand{\pvec}{\mbs{\phi}}
\newcommand{\coef}{\mbf{c}}

\begin{document}

\title{PUMA criterion = MODE criterion}
\author{Dave Zachariah, Petre Stoica and Magnus Jansson\thanks{This work has been partly supported by the Swedish Research Council
(VR) under contracts 621-2014-5874 and 2015-05484.}}

\maketitle

\begin{abstract}
We show that the recently proposed (enhanced) PUMA estimator for array
processing minimizes the same criterion function as the well-established MODE estimator. (PUMA = principal-singular-vector utilization for modal analysis, MODE = method of direction
estimation.)
\end{abstract}

\section{Problem formulation}

The standard signal model in array processing is
\begin{equation}
\y(t) = \A(\pvec) \s(t) + \n(t) \in \mathbb{C}^{m}
\end{equation}
where $\pvec = [\phi_1 \: \cdots \: \phi_r]^\top$ parameterizes the unknown directions of arrival from $r < m$ far-field sources, $\s(t)$ is a vector of unknown source signals, $\n(t)$ is a noise term, and $\A(\cdot)$ is a known
function describing the array response \cite{Stoica&Moses2005_spectral,vanTrees2004_array}. The covariance matrix of the
received signals is
\begin{equation}
\ycov = \A\scov \A^* + \sigma^2 \I_m,
\end{equation}
where $\scov$ and $\sigma^2 \I_m$ are the signal and noise covariances, respectively. The data is assumed to be circular Gaussian.

Given $T$ independent snapshots $\{ \y(t) \}^T_{t=1}$, the maximum
likelihood (ML) estimate of $\pvec$ is given by
\begin{equation}\label{eq:ML}
\what{\pvec} = \argmin_{\pvec} \; \tr\left\{  \mbs{\Pi}^\perp_{\A}
  \what{\ycov}  \right\},
\end{equation}
where 
\begin{equation*}
\what{\ycov} = \frac{1}{T}\sum^{T}_{t=1} \y(t) \y^*(t)
\end{equation*}
denotes the sample covariance matrix and $\mbs{\Pi}^\perp_{\A}$ is the
orthogonal projector onto $\mathcal{R}(\A)^\perp$ and is a nonlinear function
of $\pvec$. The nonconvex problem in \eqref{eq:ML} can be viewed as 
fitting the signal subspace spanned by $\A$ to the data, and it can be
tackled using numerical search techniques.

When considering uniform linear arrays, the columns of $\A$ have a
Vandermonde structure:
\begin{equation*}
\A = \begin{bmatrix}
1               & 1                & \cdots & 1 \\
e^{j \phi_1} & e^{j \phi_2}  & \cdots & e^{j \phi_r} \\
\vdots      & \vdots        &             & \vdots \\
e^{j (m-1) \phi_1} & e^{j (m-1) \phi_2}  & \cdots & e^{j (m-1)\phi_r} 
\end{bmatrix}.
\end{equation*}
 In this case we have the following orthogonal relation
\begin{equation}\label{eq:orthogonal}
\T \A = \0
\end{equation}
where
\begin{equation*}
\T = 
\begin{bmatrix}
c_0 & c_1      & \cdots & c_r & &\\
0   & \ddots & \ddots &      & \ddots & 0 \\
     &             &  c_0   &  c_1    &  \cdots & c_r
\end{bmatrix} \in \mathbb{C}^{(m-r) \times m}
\end{equation*}
is a Toeplitz matrix with coefficients $\coef = [c_0
\; c_1 \cdots \; c_r]^\top$. These coefficients also define a
polynomial with roots that lie on the unit circle,
\begin{equation*}
c_0 + c_1 z + \cdots + c_r z^r = c_0 \prod^r_{k=1} (1 - e^{-j \phi_k}
z), \quad c_0 \neq 0.
\end{equation*}
Therefore there is a direct correspondence between $\pvec$
and $\coef$ \cite{Stoica&Moses2005_spectral,vanTrees2004_array}.
As a consequence of \eqref{eq:orthogonal} the orthogonal projector can
be written as
\begin{equation*}
\mbs{\Pi}^\perp_{\A} = \mbs{\Pi}_{\T} = \T^* (\T \T^*)^{-1} \T 
\end{equation*}
which yields an equivalent problem to \eqref{eq:ML} in terms of
$\coef$:
\begin{equation}\label{eq:ML_alt}
\what{\coef} = \argmin_{\coef} \; V_{\text{ML}}(\coef),
\end{equation}
where
\begin{equation}\label{eq:V_ML}
V_{\text{ML}}(\coef) = \tr\left\{  \mbs{\Pi}_{\T}
  \what{\ycov}  \right\} = \tr\left\{ (\T \T^*)^{-1} \T 
  \what{\ycov}  \T^* \right\}.
\end{equation}
Using this alternative parameterization, tractable
minimization algorithms can be formulated. Next, we consider two
alternative estimation criteria and prove that they are equivalent.

\section{PUMA criterion equals MODE criterion}

Using the eigendecomposition, the covariance matrix can be written as
\begin{equation*}
\ycov = \mbf{U}_s \Lam \mbf{U}^*_s + \sigma^2 \mbf{U}_n \mbf{U}^*_n
\end{equation*}
where $\mathcal{R}( \mbf{U}_s ) = \mathcal{R}( \A ) $ and $\Lam =
\text{diag}(\lambda_1, \dots, \lambda_r) \succ \0$ is the matrix of eigenvalues that are larger than $\sigma^2$. Instead of
fitting the subspace to the sample covariance $\what{\ycov}$, as in
\eqref{eq:V_ML}, consider fitting to a weighted estimate of the signal
subspace \cite{Stoica&Sharman1990_mode,Viberg&Ottersten1991_wsf}: 
\begin{equation*}
\U \G \U^*,
\end{equation*}
where 
\begin{equation*}
\G  \triangleq \text{diag}\left( \frac{(\hat{\lambda}_1 - \hat{\sigma}^2)^2}{  \hat{\lambda}_1
  }, \dots,  \frac{(\hat{\lambda}_r - \hat{\sigma}^2)^2}{  \hat{\lambda}_r } \right)
\end{equation*}
and where $\{ \hat{\lambda}_i \}$ and $\hat{\sigma}^2$ are obtained from the eigendecomposition of $\what{\ycov}$.
Then the cost function in \eqref{eq:ML_alt} is replaced by 
\begin{equation*}
V_{\text{MODE}}(\coef) = \tr\left\{ (\T \T^*)^{-1} \T  \U \G \U^* \T^* \right\}.
\end{equation*}
This leads to the asymptotically efficient `\emph{m}ethod \emph{o}f \emph{d}irection
\emph{e}stimation' (\textsc{Mode})
\cite{Stoica&Sharman1990_mode}\cite[ch.~8.5]{vanTrees2004_array}. A
simple two-step algorithm was proposed in
\cite{Stoica&Sharman1990_mode} to approximate the minimum of the above
estimation criterion.


Another approach for array processing, called `\emph{p}rincipal-singular-vector \emph{u}tilization for
\emph{m}odal \emph{a}nalysis'  (\textsc{Puma}), has been recently
proposed in \cite{QianEtAl2016_puma} (see also references therein for
predecessors of that approach). It is motivated by properties of a related linear
prediction problem and based on the following fitting criterion
\begin{equation*}
V_{\text{PUMA}}(\coef) = \e^* \W \e,
\end{equation*}
where
\begin{equation*}
\begin{split}
\W & \triangleq ( \G \otimes (\T \T^*)^{-1} )
\end{split}
\end{equation*}
is a weighting matrix and $\e$ is a function of $\coef$ and the
eigenvectors in $\U$. As shown in \cite{QianEtAl2016_puma}, $\e$ can be written as $\e = \text{vec}( \T \U )$. It follows immediately that
\begin{equation*}
\begin{split}
V_{\text{PUMA}}(\coef) &= \e^* \W \e\\
&= \mvec( \T \U )^*\left( \G \otimes (\T \T^*)^{-1} \right) \mvec( \T \U ) \\
&= \mvec( \T \U )^*\mvec( (\T \T^*)^{-1} \T \U \G ) \\
&= \tr\left\{   \U^* \T^* (\T \T^*)^{-1} \T \U \G  \right\} \\
&= \tr\left\{    (\T \T^*)^{-1} \T \U \G \U^* \T^* \right\} \\
&= V_{\text{MODE}}(\coef) ,
\end{split}
\end{equation*}
where we made use of the following results
\begin{equation*}
\begin{split}
\mvec(\mbf{XYZ}) &= ( \mbf{Z}^\top \otimes \mbf{X} ) \mvec(\mbf{Y}) \\
\text{tr}\{ \mbf{X}^* \mbf{Y}  \} &= \mvec(\mbf{X})^* \mvec(\mbf{Y}) .
\end{split}
\end{equation*}
Therefore the \textsc{Puma} criterion \emph{is exactly equivalent} to
the \textsc{Mode} criterion. The algorithm proposed in
\cite{QianEtAl2016_puma} is thus an alternative technique for minimizing $V_{\text{MODE}}(\coef)$.

\section{Other variants}

A fitting criterion on a similar form as
$V_{\text{PUMA}}(\coef)$ was proposed in
\cite{JanssonEtAl1998_weighted} and shown to reduce to
$V_{\text{MODE}}(\coef)$ in a special case. Alternative
minimization techniques are also discussed therein, see also
\cite[ch.~8]{vanTrees2004_array}. See
e.g. \cite{Stoica&Jansson1997_forward,KristenssonEtAl1999_modified} for additional variations of $V_{\text{MODE}}(\coef)$.  

In scenarios with low signal-to-noise ratio or small sample size $T$, subspace-fitting methods such as
\textsc{Mode} may suffer from a threshold breakdown effect due to `subspace swaps' \cite{HawkesEtAl2001_performance,JohnsonEtAl2008_music}. To 
reduce the risk that the signal subspace is fitted to noise in these
cases, a modification was proposed in
\cite{Gershman&Stoica1999_modex} consisting of using $p < m - r$ extra coefficients
in $\coef$. Then after computing the corresponding directions of
arrival, all possible subsets of $r$ directions are compared using the
maximum likelihood criterion and the best subset is chosen as the
estimate. This method is called the \textsc{ModeX} in
\cite{Gershman&Stoica1999_modex} and its principle is exactly what is
used in \cite{QianEtAl2016_puma} to propose the Enhanced-\textsc{Puma}.

Interestingly, while both papers \cite{Stoica&Sharman1990_mode} and
\cite{Gershman&Stoica1999_modex} are referenced in
\cite{QianEtAl2016_puma}, the equivalence (as shown above) of the
\textsc{Puma} estimation criterion proposed there to \textsc{Mode}
\cite{Stoica&Sharman1990_mode}  and \textsc{ModeX} estimation criteria \cite{Gershman&Stoica1999_modex} was missed in \cite{QianEtAl2016_puma}.


\bibliographystyle{ieeetr}
\bibliography{refs_pumamode}
\end{document}